\documentclass[a4paper]{jpconf}
\usepackage{graphicx}
\usepackage{lineno}
\usepackage{cite}
\let\citedash\relax
\makeatletter \providecommand{\citedash}{\hbox{$\sim$}\penalty\@m}
\makeatother
\begin{document}
\title{Energy Dependence of Moments of Net-Proton, Net-Kaon, and Net-Charge Multiplicity Distributions at STAR}

\author{Ji Xu (for the STAR Collaboration)}

\address{ Key Laboratory of Quark \& Lepton Physics (MOE) and Institute of Particle Physics, Central China Normal University, Wuhan 430079, China

   Lawrence Berkeley National Laboratory, Berkeley, CA 94720, US}

\ead{jixu@lbl.gov}

\begin{abstract}
One of the main goals of the RHIC Beam Energy Scan (BES) program is to study the QCD phase structure, which includes the search for the QCD critical point, over a wide range of chemical potential ($\mu_{B}$). Theoretical calculations predict that fluctuations of conserved quantities, such as baryon number (B), charge (Q), and strangeness (S), are sensitive to the correlation length of the dynamical system. Experimentally, higher moments of multiplicity distributions have been utilized to search for the QCD critical point in heavy-ion collisions. In this paper, we report recent efficiency-corrected cumulants and cumulants ratios of the net-proton, net-kaon, and net-charge multiplicity distributions in Au+Au collisions at $\sqrt{s_{NN}}$ = 7.7, 11.5, 14.5, 19.6, 27, 39, 62.4, and 200 GeV collected in the years 2010, 2011, and 2014 with STAR at RHIC. The centrality and energy dependence of the cumulants up to the fourth order, as well as their ratios, are presented. Furthermore, the comparisons with baseline calculations (Poisson) and non-critical-point models (UrQMD) will also be discussed.
\end{abstract}

\section{Introduction}
Fluctuations of conserved quantities, such as net-baryon ($\Delta{N_{B}}$), net-charge ($\Delta{N_{Q}}$) and net-strangeness ($\Delta{N_{S}}$),  have long been predicted to be sensitive to the QCD phase transition and the QCD critical point~\cite{SearchCP1,SearchCP2,SearchCP4}. Experimentally, one can measure various order moments (variance ($\sigma^2$), skewness ($S$), and kurtosis ($\kappa$)) of the conserved quantity distributions in heavy-ion collisions. These moments are sensitive to the correlation length ($\xi$) of the hot dense matter created in the initial collision and are also connected to the thermodynamic susceptibilities computed with Lattice QCD~\cite{LQCD1,LQCD2,LQCD3} and in the Hadron Resonance Gas (HRG) model~\cite{HRG1,HRG2,HRG3,HRG4}. It has been proposed that variance, skewness, and kurtosis is related to different powers of the correlation length as $\xi^2$, $\xi^{4.5}$, and $\xi^7$~\cite{SearchCP1}, respectively. Furthermore, the $n^{th}$ order susceptibility $\chi ^{(n)}$ are related to cumulant as ${\chi ^{(n)}} = C_n^{}/V{T^3}$~\cite{SearchCP3}, where $V,T$ are the volume and temperature of the system, and $C_{n}$ is the $n^{th}$ order cumulant of a multiplicity distribution. In order to compare with theoretical calculations, cumulant ratios ($S\sigma=C_{3}/C_{2}$, $\kappa\sigma^2=C_{4}/C_{2}$) are constructed to cancel the volume ($V{T^3}$). Thus, those cumulant ratios are also directly related to the ratios of various order susceptibilities as $C_{4}/C_{2} = \chi_{i}^{(4)}/\chi_{i}^{(2)}$ and $C_{3}/C_{2} = \chi_{i}^{(3)}/\chi_{i}^{(2)}$, where $i$ indicates the conserved quantity. 
 
Experimentally, it is very hard to measure the net-baryon ($\Delta{N_{B}}$) and the net-strangeness ($\Delta{N_{S}}$) distributions, so we use net-proton ($\Delta{N_{P}}$) and net-kaon ($\Delta{N_{K}}$) as proxies respectively.
These have been widely studied experimentally and theoretically~\cite{HRG1,SearchCP1,SearchCP2,SearchCP3,SearchCP4,SearchCP5,SearchCP6,xflQM2015}. 
In this work, we report recent efficiency-corrected cumulants and cumulant ratios of the net-proton ($\Delta{N_{P}}$), net-kaon ($\Delta{N_{K}}$) and net-charge ($\Delta{N_{Q}}$) multiplicity distributions in Au+Au collisions at $\sqrt{s_{NN}}$ = 7.7, 11.5, 14.5, 19.6, 27, 39, 62.4, and 200 GeV collected in 2010, 2011, and 2014 by STAR at RHIC.

\section{Analysis Details}
The STAR (Solenoidal Tracker At RHIC) detector at BNL has a large uniform acceptance at mid-rapidity and excellent particle identification capabilities.
Energy loss ($dE/dx$) in the Time Projection Chamber and mass-squared ($m^{2}$) from the Time-Of-Flight detector are used to identify protons and kaons.
The protons and anti-protons are obtained at mid-rapidity ($|y|$$<$$0.5$) and within the transverse momentum range 0.4$<$$p_{T}$$<$2.0 GeV/c. The $K^{+}$ ($K^{-}$) for net-kaon are measured at mid-rapidity ($|y|$$<$0.5) within the transverse momentum 0.2$<$$p_{T}$$<$1.6 GeV/c. The charged particles in the net-charge fluctuations are measured at pseudo-rapidity range $|\eta|$$<$0.5 and within the transverse momentum 0.2$<$$p_{T}$$<$2.0 GeV/c.

For the centrality selection, the uncorrected multiplicity distribution of primary charged particles within 0.5$<$$|\eta|$$< $1.0 has been used to determine the centrality for the $\Delta{N_{Q}}$ measurement, while the uncorrected multiplicity distribution of primary charged particles, excluding kaons/anti-kaons (protons/anti-protons) within $|\eta|$$<$1.0 has been used for $\Delta{N_{K}}$ ($\Delta{N_{P}}$) to avoid auto-correlations.
The results are corrected for a finite centrality bin width~\cite{technique}.

\begin{figure*}
\hspace{0cm}
\includegraphics[width=6.in]{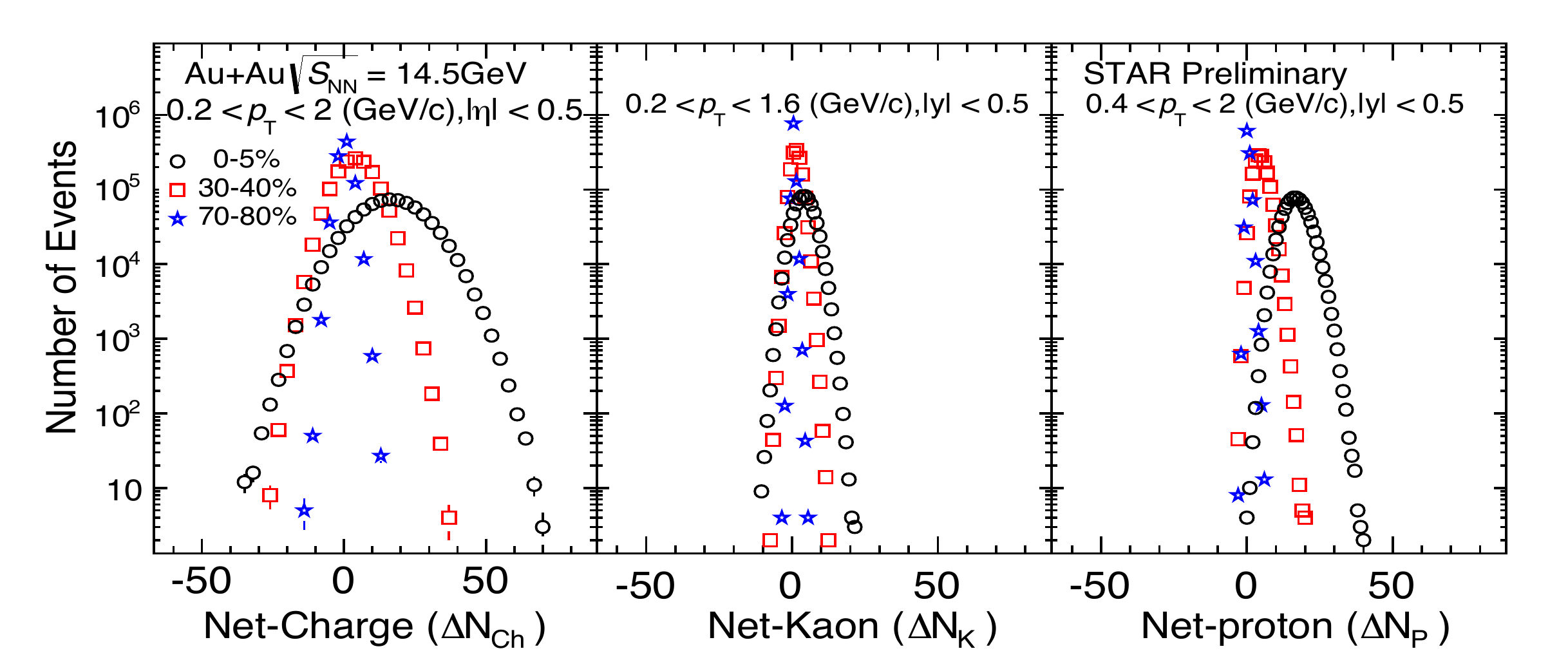}
\caption{Uncorrected raw event-by-event net-particle multiplicity distributions for Au+Au collisions at $\sqrt{s_{NN}}$ = 14.5 GeV for $\Delta{N_{Q}}$ (left panel), $\Delta{N_{K}}$ (middle panel), and $\Delta{N_{P}}$ (right panel) for 0-5\% top central (black circles), 30-40\% central (red squares), and 70-80\% peripheral collisions (blue stars).}
\label{distributions}
\end{figure*}

Figure~\ref{distributions} shows the uncorrected event-by-event net-particle multiplicity distributions for Au+Au collisions at $\sqrt{s_{NN}}$ = 14.5 GeV for $\Delta{N_{Q}}$, $\Delta{N_{K}}$ and $\Delta{N_{P}}$ in three centrality intervals. We observed that the net-charge multiplicity distributions have the largest standard deviation, $\sigma$, compared with the net-proton and net-kaon results. This can lead to a large statistical error for the net-charge measurement. Detailed discussions about the efficiency correction and error estimation can be found in~\cite{technique,error}.

\section{Results and Discussion} 

\begin{figure*}
\includegraphics[width=2.95in]{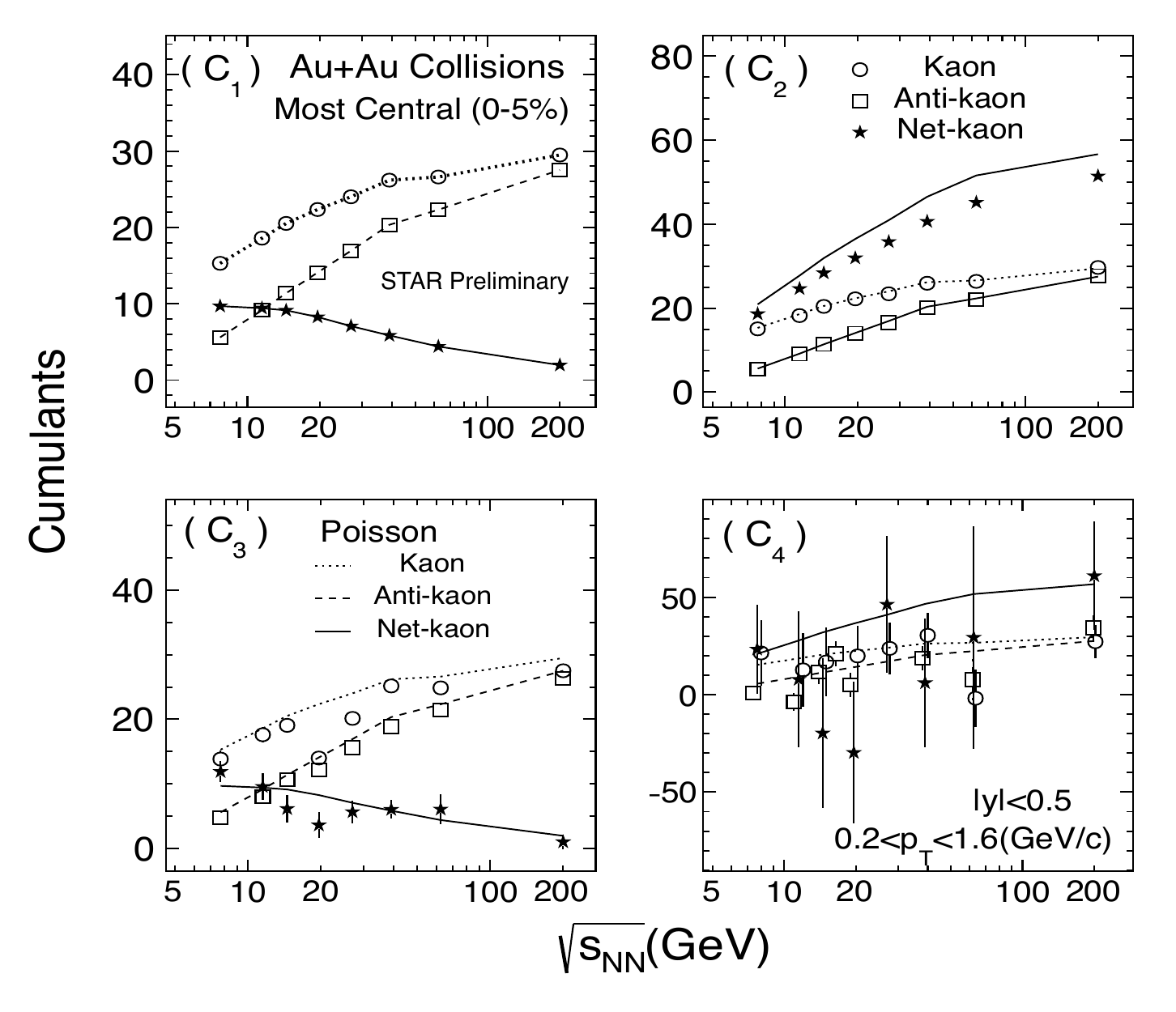}
\includegraphics[width=2.9in]{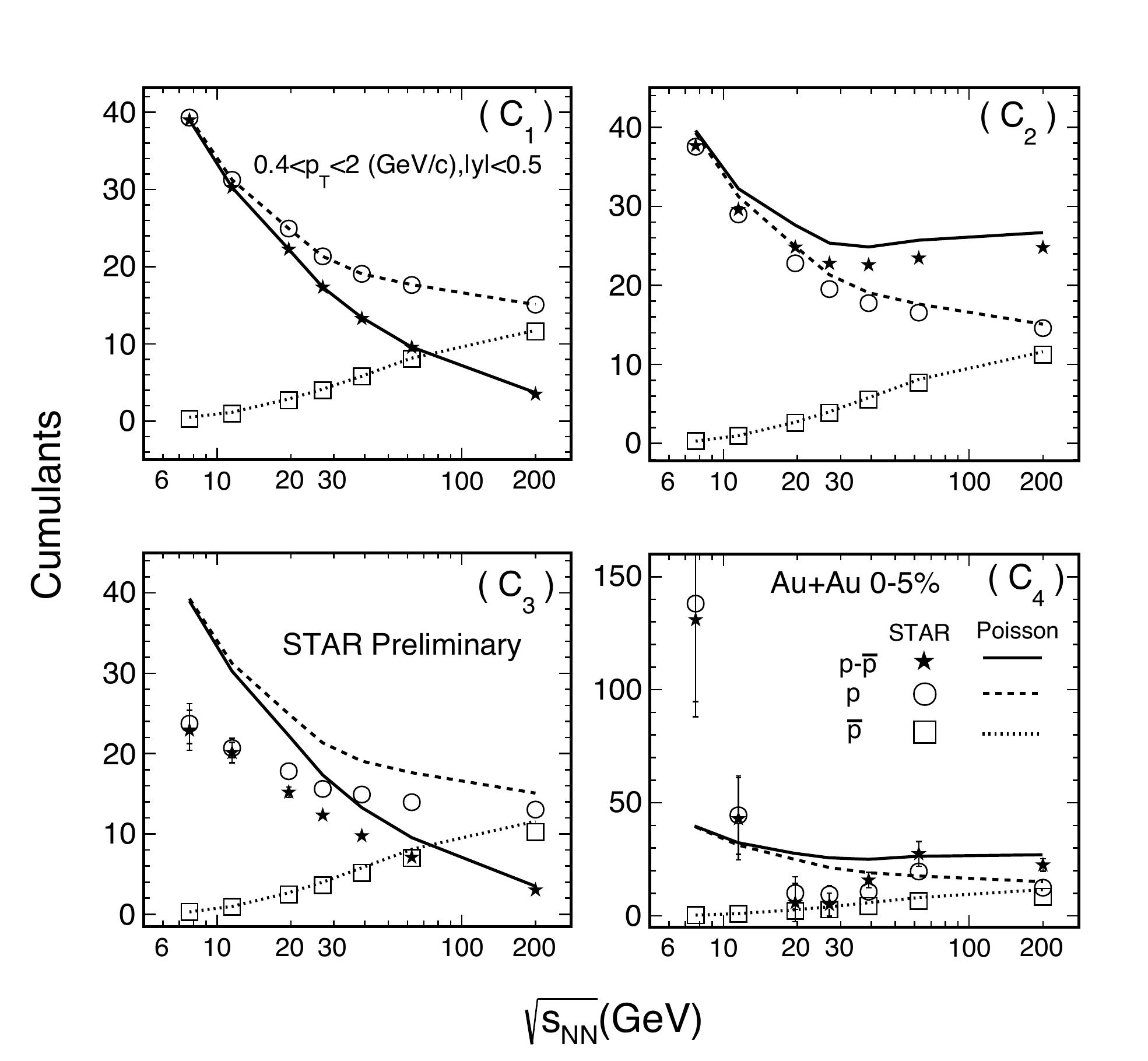}
\caption{Left panel: Energy dependence of cumulants ($C_{1}$, $C_{2}$, $C_{3}$, and $C_{4}$) for kaon, anti-kaon, and net-kaon multiplicity distributions. Right panel: Energy dependence of cumulants ($C_{1}$, $C_{2}$, $C_{3}$, and $C_{4}$,) for proton, anti-proton, and net-proton multiplicity distributions.}
\label{cumu_egy}
\end{figure*}

\begin{figure*}
\includegraphics[width=2in]{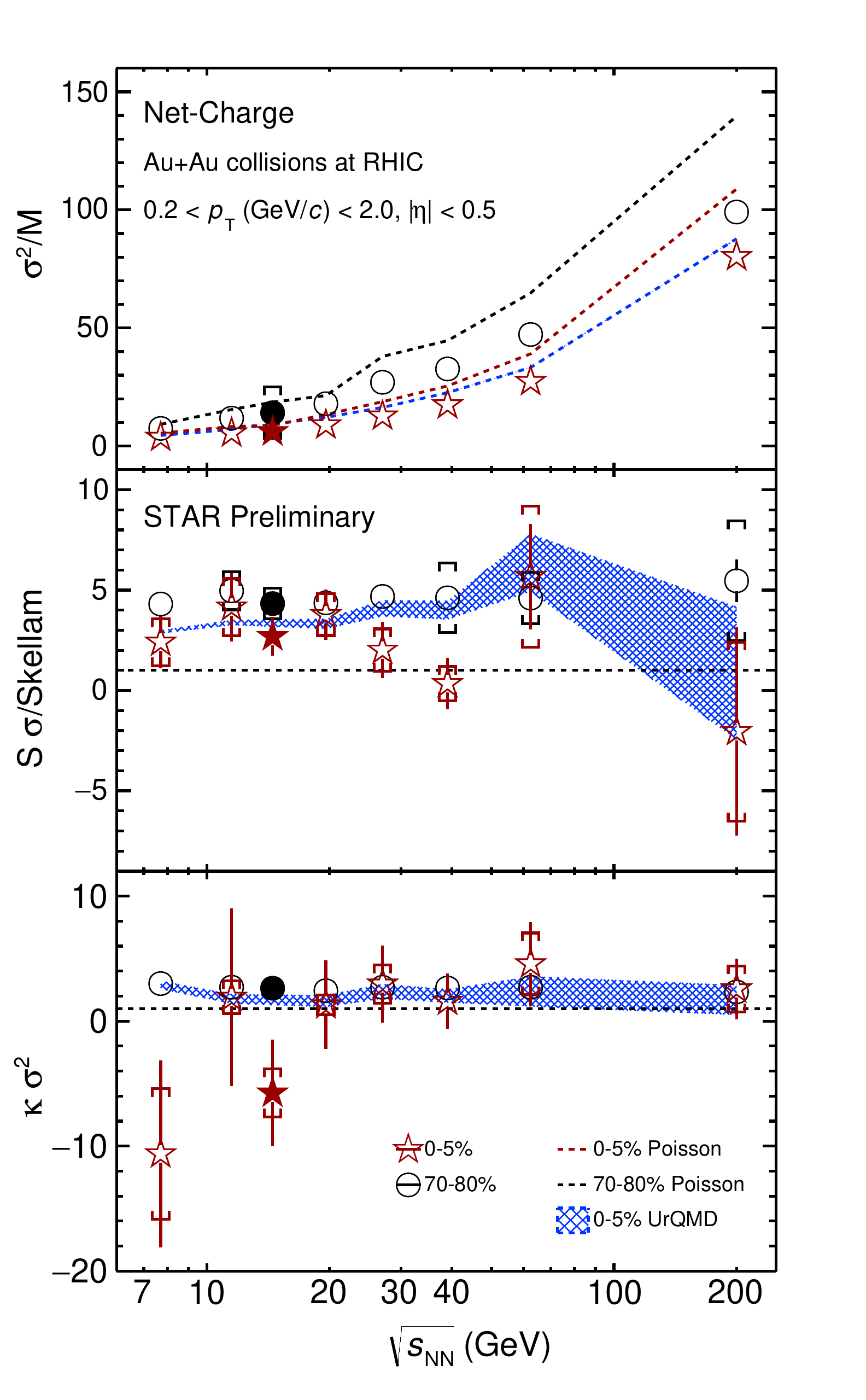}
\includegraphics[width=2in]{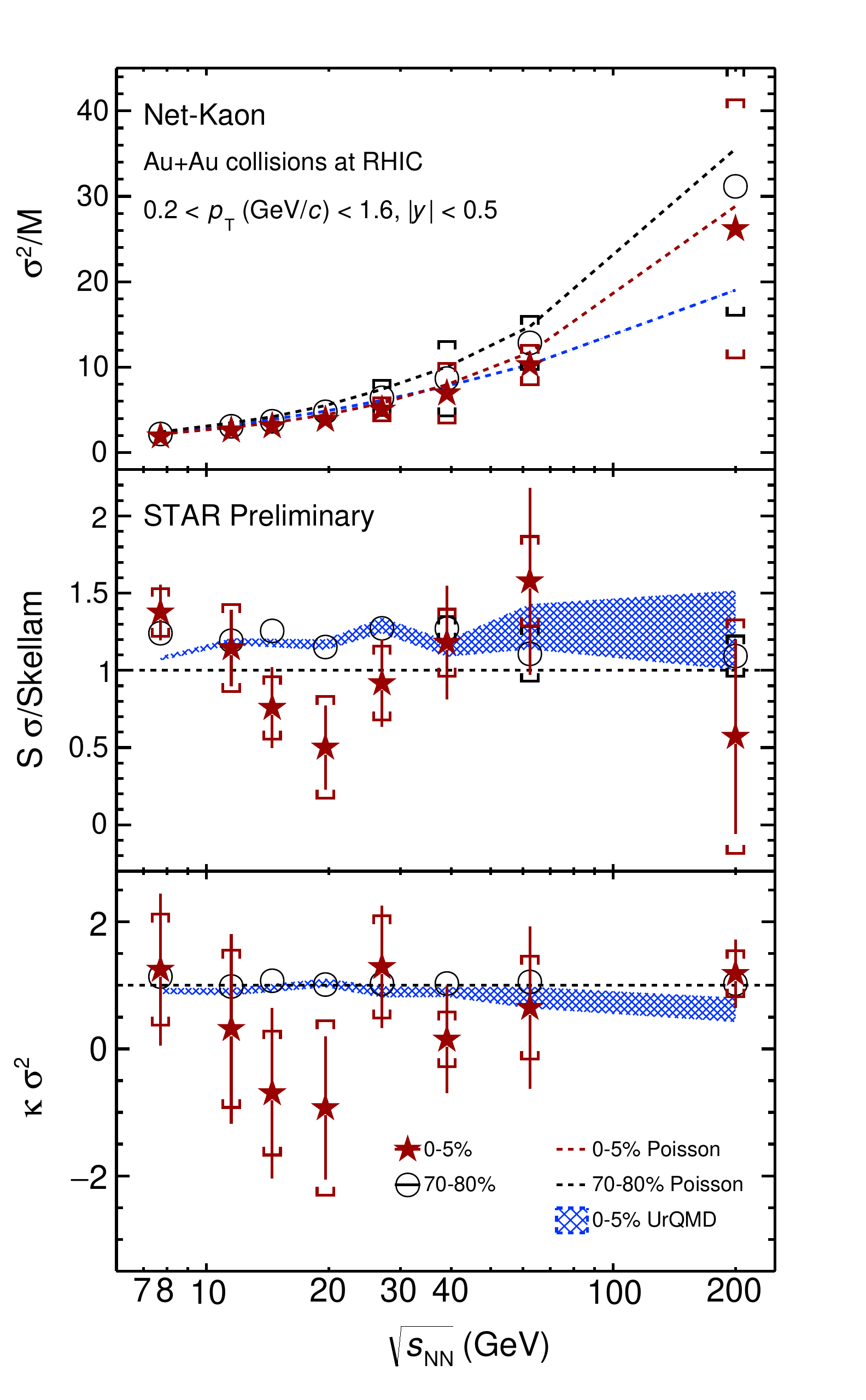}
\includegraphics[width=2in]{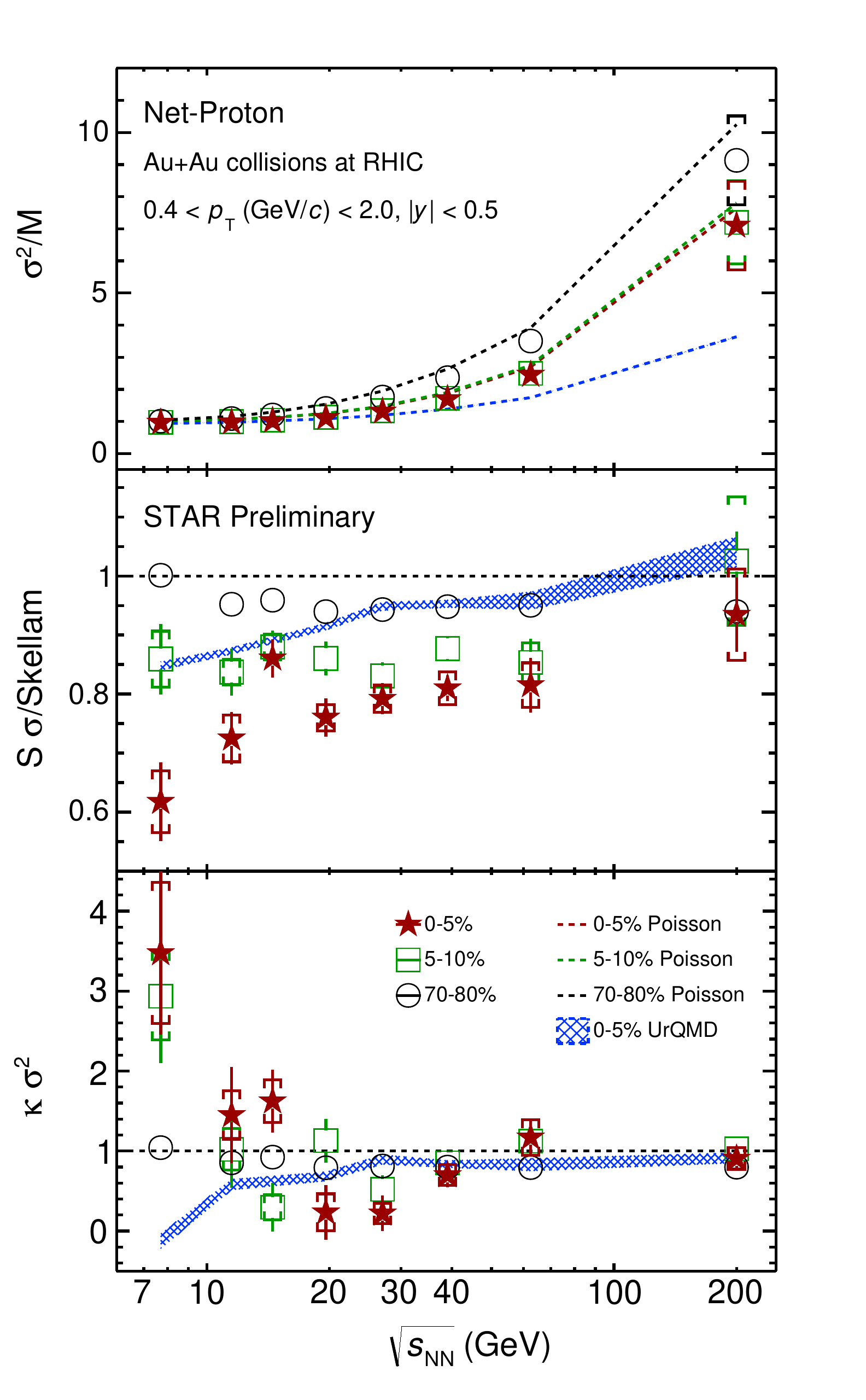}
\caption{ Energy dependence of cumulant ratios ($\sigma^{2}/M$, $S\sigma/$Skellam, $\kappa\sigma^2$) of net-charge, net-kaon, and net-proton multiplicity distributions for top 0-5\% central (red stars), 5-10\% central (green squares), and 70-80\% peripheral (black circles) collisions. The Poisson expectations are denoted as dotted lines and UrQMD calculations are shown as blue bands.} 
\label{egyDependence}
\end{figure*}

All results have been corrected for finite detector reconstruction efficiencies, with the statistical uncertainties calculated based on the Delta Theorem~\cite{error}.
The left panel of Fig.~\ref{cumu_egy} shows the energy dependence of cumulants ($C_{1}$, $C_{2}$, $C_{3}$, and $C_{4}$) for kaon, anti-kaon, and net-kaon multiplicity distributions. 
Dashed lines indicate the Poisson expectations. In general, the cumulants for net-kaon, kaon and anti-kaon are consistent with the Poisson baseline within uncertainties.
The right panel of Fig.~\ref{cumu_egy} shows the energy dependence of cumulants ($C_{1}$, $C_{2}$, $C_{3}$, and $C_{4}$) for proton, anti-proton, and net-proton multiplicity distributions. One can see that the higher the order of the cumulant, the larger the deviations from the Poisson expectation for the net-protons and protons.

The energy dependence of the volume independent cumulant ratios ($\sigma^{2}/M$, $S\sigma/$Skellam, $\kappa\sigma^2$) for net-proton, net-kaon, and net-charge multiplicity distributions in Au+Au collisions are presented in Figure~\ref{egyDependence}.
The solid circles on the left figure represent the results of $\sqrt{s_{NN}}$ = 14.5 GeV obtained from Run 14 at RHIC, added to the published net-charge results~\cite{netcharge} (open symbols).
The blue bands give the predictions from the UrQMD calculations which do not include critical physics processes.
No significant deviation from the Poisson expectation (dashed lines in the figure) is observed within the uncertainties for net-charge and net-kaon cumulant ratios.
Both the data and the UrQMD simulation show no energy dependence for $S\sigma/$Skellam and $\kappa\sigma^2$ in net-charge and net-kaon measurement. 

However, a deviation from the Poisson expectation, as well as from the UrQMD calculations is observed for $S\sigma/Skellam$ and $\kappa\sigma^2$ of the net-proton distributions. 
A non-monotonic behavior of the net-proton $\kappa\sigma^2$ can be seen in top 0-5\% central collisions; the value of $\kappa\sigma^2$ go down below unity and then rise up from high to low collision energies. UrQMD calculations show suppression at lower energies due to baryon number conservation.

The upcoming RHIC BES II in 2019-2020, will include an upgraded STAR detector. An i(nner)TPC and Endcap TOF upgrade will enlarge the phase-space up to $|\eta|$$<$1.5 and down to $p_{T}$ = 60 MeV/c. The Event-Plane Detector at forward rapidities will allow for a better centrality estimation, suppressing auto-correlations. 

\section{Summary}
In this paper, we present the energy dependence on cumulant ratios from $\sqrt{s_{NN}}$ = 7.7 to 200 GeV Au+Au collisions for net-kaon, net-charge
and net-proton multiplicity distributions. Within uncertainties (statistical and systematic), net-charge and net-kaon results follow the Poisson
expectation, while a non-monotonic behavior is observed in the energy dependence for the net-proton $\kappa\sigma^2$. 
The energy dependence of cumulants ($C_{1}$ , $C_{2}$, $C_{3}$, and $C_{4}$) for kaon and proton multiplicity distributions have also been shown, and 
in general, the cumulants are consistent with the Poisson baseline within uncertainties for kaon, anti-kaon, and net-kaon. We can see that for net-proton and proton distributions, the deviations from Poisson expectations are larger for the higher order cumulants. The RHIC BES II will bring a larger event sample and a wider phase-space to enhance the search for the QCD critical point. 

\section*{Acknowledgments}
The work was supported in part by the MoST of China 973-Project No.2015CB856901, NSFC
under grant No. 11575069, 11221504.


\end{document}